\documentclass[10pt]{iopart}
\usepackage{graphicx}
\usepackage{color}

\begin{document}

\title{Quantitative Characterization of the Microstructure
and Transport Properties of Biopolymer Networks}

\author{Yang Jiao$^1$, Salvatore Torquato$^{1,2,3,4,5}$}

\address{$^1$ Physical Science in Oncology Center,
Princeton Institute for the Science and Technology of Materials,
Princeton University,
Princeton, NJ 08544, USA \\

$^2$  Department of Chemistry,
Princeton University,
Princeton, NJ 08544, USA  \\

$^3$  Department of Physics,
Princeton University,
Princeton, NJ 08544, USA  \\

$^4$ Princeton Center for Theoretical Science,
Princeton University,
Princeton, NJ 08544, USA \\

$^5$ Program in Applied and Computational Mathematics,
Princeton University,
Princeton, NJ 08544, USA \\
}

\vspace{1in} \noindent \textbf{Corresponding author contact information}:
\begin{tabbing}
 \hspace{0.25in}
 \= Salvatore Torquato \\
 \>  Tel.: 609-258-3341 \\
 \>  Fax: 609-258-6746 \\
 \>  E-mail: torquato$@$princeton.edu
\end{tabbing}

\noindent \textbf{Short title}: Microstructure and transport properties of collagen \newline

\noindent \textbf{Classification numbers}: 87.15.rp, 87.85.jc

\newpage

\begin{abstract}

Biopolymer networks are of fundamental importance to many
biological processes in normal and tumorous tissues. In this
paper, we employ the panoply of theoretical and simulation
techniques developed for characterizing heterogeneous materials to
quantify the microstructure and effective diffusive transport
properties (diffusion coefficient $D_e$ and mean survival time
$\tau$) of collagen type I networks at various collagen
concentrations. In particular, we compute the pore-size
probability density function $P(\delta)$ for the networks and
present a variety of analytical estimates of the effective
diffusion coefficient $D_e$ for finite-sized diffusing particles,
including the low-density approximation, the Ogston approximation,
and the Torquato approximation. The Hashin-Strikman upper bound on
the effective diffusion coefficient $D_e$ and the pore-size lower
bound on the mean survival time $\tau$  are used as benchmarks to
test our analytical approximations and numerical results.
Moreover, we generalize the efficient first-passage-time
techniques for Brownian-motion simulations in suspensions of
spheres to the case of fiber networks and compute the associated
effective diffusion coefficient $D_e$ as well as the mean survival
time $\tau$, which is related to nuclear magnetic resonance (NMR)
relaxation times. Our numerical results for $D_e$ are in excellent
agreement with analytical results for simple network
microstructures, such as periodic arrays of parallel cylinders.
Specifically, the Torquato approximation provides the most
accurate estimates of $D_e$ for all collagen concentrations among
all of the analytical approximations we consider. We formulate a
universal curve for $\tau$ for the networks at different collagen
concentrations, extending the work of Yeong and Torquato [J. Chem.
Phys. {\bf 106}, 8814 (1997)]. We apply rigorous
cross-property relations to estimate the effective bulk modulus 
of collagen networks from a knowledge of the effective diffusion coefficient 
computed here. The use of cross-property relations to link other 
physical properties to the transport properties of collagen networks is also
discussed.

\end{abstract}

\vspace{2pc}
\maketitle

\section{Introduction}


Biopolymer networks, such as the cross-linked bundles (or fibers)
of collagen and fibrin in the extracellular matrix (ECM), provide
mechanical support for cells and serve as the media for the
transmission of many biomechanical/biochemical cues that regulate
cell motility, proliferation, differentiation and apoptosis
\cite{kuntz97, lo00, bisch03, grin03}. The diffusion and
absorption of various macromolecules in biopolymer networks are of
crucial importance to the regulation and metabolism of normal
organs and to the delivery of drugs in tumor tissues
\cite{comper96, jana08}. Such biological processes are largely determined by
the composition and microstructure of the network, especially the
complex pore space between the fibers \cite{jain87, yang10}.
Thus, knowledge of the effective transport and mechanical properties
of biopolymer networks is crucial in order to understand quantitatively
the aforementioned biological processes.


The microstructure of biopolymer networks and their associated
transport properties have been investigated by many researchers.
For example, Ogston et al. \cite{ogston58, ogston73} introduced
the idea of ``influence cylinders'' associated with each fiber in
the system, which enables one to obtain the probability
distribution of spherical pores with different sizes $\delta$,
i.e., the {\it pore-size probability density function} $P(\delta)$
(also referred to as the {\it pore-size distribution function} in
literature; see the definition in Sec. 2). For polymer networks
composed of very long and stiff fibers, Ogston derived an
analytical expression of $P(\delta)$, which depends on the volume
fraction and the characteristic diameter of the fibers
\cite{ogston58}. Other approaches used to ascertain pore-size
statistics include Fourier analyses of three-dimensional (3D)
confocal microscopy images \cite{taka03}, or statistical analysis
of nearest points on collagen fibers obtained from
confocal-microscopic image stacks \cite{kauf05}. Recently, a
method based on the direct analysis of the entire 3D real-space
network geometry from high-contrast confocal microscopy data has
been developed \cite{mick08, lind10}.
Specifically, Lindstr{\"{o}}m et al. \cite{lind10} represented
the collagen fibers in the networks as thinned skeletal center
lines and the cross-links are represented as nodes, which can be
thought of as the ``graph'' representation of a biopolymer network. 
These authors also employed an inverse
reconstruction method to characterize the microstructure of the
collagen networks and investigated the mechanical properties of
the networks using finite-element analysis.

The determination of the effective diffusion coefficient $D_e$ for
polymer networks dates back to the pioneering work of Johansson,
L{\"{o}}froth and coworkers \cite{joha91a, joha93a, joha91b,
joha91c, joha93b}. Johansson et al. experimentally studied the
diffusion of small monodisperse polyethylene glycols
\cite{joha91a} and nonionic micelles \cite{joha93a} in polymer
systems and accurately measured the long-time-limit self-diffusion
coefficient (i.e., $D_e$) using a tracer technique \cite{joha91b}.
By considering the local diffusion of a particle around a single
fiber, Johansson et al. \cite{joha91c} derived an analytical
approximation of $D_e$ which incorporates the microstructural
information of the pore space. Since their approach was based on
the key concept of the distribution of ``influence cylinders''
introduced by Ogston, the approximation of $D_e$ is henceforth
referred to as the Ogston approximation. To test the predictive
capacity of their theory, Johansson and L{\"{o}}froth
\cite{joha93b} carried out hard-sphere Brownian-motion
simulations, in which the diffusing particles were hard spheres
and the fibers were considered to hinder the diffusion of the
particles. Although hydrodynamic effects were not taken into
account in their simulations, the results were shown to be in
excellent agreement with experimental data and theoretical
predictions for a wide range of particle sizes  \cite{joha93b}.
Recently, the effects of the anisotropy of fiber orientations
\cite{ledd06, erik08} and of the hydrodynamic interactions between
the particle and fibers \cite{styl10} on the effective diffusion
coefficient have also been investigated. We note
that by mathematical analogy, the problem of macromolecular
diffusion in the pore space exterior to the collagen fibers is
equivalent to the electrical or thermal conduction problem in the
pore space with perfectly insulating fibers \cite{SalBook}.


Very recently, it has been suggested that the powerful
theoretical and simulation techniques developed for characterizing
the microstructure and effective properties of random
heterogeneous materials \cite{SalBook, sahimi1, sahimi2, zohdi} could be fruitfully employed to model
complex biological systems, especially malignant tumors and the
associated host microenvironment \cite{Sal11}. This idea has led
to fruitful applications in the understanding of the spatial
organizations of abnormal cells in brain tumors \cite{Jiao11}.



In this paper, we further explore these techniques from the theory 
of heterogeneous materials by investigating the
microstructure and transport properties of collagen type I
networks (i.e., the most abundant collagen of the human body found
in tissue and bones, and therefore, called ``type I''; see Fig. 1).
There exist analytical expressions that relate the effective transport
and mechanical properties of general heterogeneous materials to
their microstructure via a variety of $n$-point correlation
functions; see Ref.~\cite{SalBook} and references therein. This
formalism has led to the evaluation of effective transport and mechanical properties for a
variety of classes of microstructures, including dispersions of
penetrable \cite{A1, A2, A3}, impenetrable spheres \cite{B1, B2,
B3}, oriented fibers \cite{C1, C2} and ellipsoid suspensions
\cite{D, SalBook}, fluid-saturated rock \cite{E}, and
interpenetrating ceramic-metal composites \cite{F}.

In the case of collagen-like networks, we calculate for the first
time a variety of structural descriptors as well as the associated
transport properties, such as the effective diffusion coefficient
$D_e$ and mean survival time $\tau$ of a Brownian particle
assuming that the fiber interface is perfectly absorbing (i.e.,
the average time that a Brownian particle spends in the solvent
before it gets trapped by the fibers). 
The mean survival time is related to the nuclear magnetic 
resonance (NMR) relaxation times as discussed below.
The latter transport
property is intimately related to the pore statistics
\cite{SalBook, ave91}. We also employ a variety of approximation
schemes for the effective diffusion coefficient $D_e$, including
the low-density approximation \cite{SalBook}, Ogston approximation
\cite{joha91c} and the Torquato approximation based on the
perturbation (phase-property contrast) expansion of $D_e$
\cite{Sal85, SalBook}. These approximations incorporate different
levels of microstructural information in terms of various
lower-order correlation functions that statistically describe the
network microstructure. The Hashin-Strikman upper bound on the
effective diffusion coefficient $D_e$ \cite{hs62} and the
pore-size lower bound on the mean survival time $\tau$
\cite{ave91, prager63}  are used as benchmarks to test our
analytical approximations as well as numerical simulations.
Specifically, we generalize the efficient first-passage-time
techniques for Brownian-motion simulations in suspensions of
spheres \cite{Sal89APL, kimall, kimall2, kimall3, kimJCP} to the
case of fiber networks and compute the associated effective
diffusion coefficient $D_e$ and mean survival time $\tau$. Our
numerical results of $D_e$ are in excellent agreement for the
analytical results of simple network microstructures such as
periodic arrays of parallel cylinders. Moreover, we show that the
Torquato approximation provides the most accurate estimate of
$D_e$ for all concentrations among the employed approximation
schemes. We also formulate a universal curve for $\tau$ for
different networks at different collagen concentrations, i.e., we
devise a way to scale $\tau$ in such a way that the scaled data
for different collagen networks collapse onto a single curve.
Rigorous cross-property relations \cite{SalBook} are applied to
estimate the effective bulk modulus of collagen networks from a knowledge
of the computed effective diffusion coefficient.


The rest of the paper is organized as follows: In Sec. 2, we
define the statistical descriptors that will be used to
characterize the network microstructures. In Sec. 3, we provide
analytical approximations and rigorous bounds for the effective
properties $D_e$ and $\tau$. In addition, we discuss the
first-passage-time simulation techniques for network structures in
detail. In Sec. 4, we present the analytical and numerical results
of the correlation functions and the effective transport
properties. In Sec. 5, we estimate the effective
bulk modulus of collagen networks using the cross-property
relations from the computed effective diffusion coefficients of
the networks. In Sec. 6, we make concluding remarks, including 
the use of cross-property relations to link other 
physical properties to the transport properties of collagen networks.


\section{Network Microstructure Characterization}

A collagen network can be considered to be a two-phase heterogeneous
material composed of a fiber phase and solvent phase (i.e.,
the pore space), which is exterior to the fibers. A important feature of
such a network microstructure is that both phases percolate across the system,
i.e., there is a continuous path between any two point of the
phase of interest that is entirely in the phase of interest, even
when the volume fraction of the fiber phase (fraction of space
covered by the fibers) is very low. In this section, we will
introduce various statistical microstructural descriptors for a
general two-phase material, including the $n$-point correlation
functions $S_n$ and the pore-size probability density function
$P(\delta)$.

\subsection{$n$-Point Correlation Functions}

Consider a two-phase heterogeneous material in which each phase
has volume fraction $\phi_i$ ($i=1,~2$), it is customary to
introduce the indicator function ${\cal I}^{(i)}({\bf x})$ defined as
\begin{equation}
\label{eq101}
{\cal I}^{(i)}({\bf x}) = \left\{
{\begin{array}{*{20}c}
{1, \quad\quad {\bf x} \in {\cal V}_i,}\\
{0, \quad\quad {\bf x} \in \bar{{\cal V}_i},}
\end{array} }\right.
\end{equation}
\noindent where ${\cal V}_i$ is the region occupied by phase $i$ and
$\bar{{\cal V}_i}$ is the region occupied by the other phase. The
statistical characterization of the spatial variations of the
material involves the calculation of the standard $n$-point
correlation functions:
\begin{equation}
\label{eq102}
S^{(i)}_n({\bf x}_1,{\bf x}_2,\cdots,{\bf x}_n) =
\left\langle{{\cal I}^{(i)}({\bf x}_1){\cal I}^{(i)}({\bf x}_2)\cdots {\cal I}^{(i)}({\bf x}_n) }\right\rangle,
\end{equation}
\noindent where the angular brackets
$\left\langle{\cdots}\right\rangle$ denote an ensemble average.
The quantity $S_n^{(i)}({\bf x}_1, {\bf x}_2, \ldots, {\bf x}_n)$ also gives
the probability of finding $n$ points positioned at ${\bf x}_1,
{\bf x}_2, \ldots, {\bf x}_n$ all in phase $i$.

For \textit{statistically homogeneous} materials, the $n$-point
correlation function depends not on the absolute positions but on
their relative displacements, i.e.,
\begin{equation}
\label{eq1002}
S^{(i)}_n({\bf x}_1,{\bf x}_2,\cdots,{\bf x}_n) = S^{(i)}_n({\bf x}_{12},\cdots,{\bf x}_{1n}),
\end{equation}
\noindent for all $n \ge 1$, where ${\bf x}_{ij}={\bf x}_j-{\bf
x}_i$. Thus, there is no preferred origin in the system, which in
Eq.~(\ref{eq1002}) we have chosen to be the point ${\bf x}_1$. In
particular, the one-point correlation function is a constant
everywhere, namely, it is equal to the volume fraction $\phi_i$ of phase $i$,
i.e.,
\begin{equation}
\label{eq103}
S^{(i)}_1 = \left\langle{{\cal I}^{(i)}({\bf x})}\right\rangle = \phi_i,
\end{equation}
\noindent which is the probability that a randomly chosen point
in the material belongs to phase $i$. For \textit{statistically
isotropic} materials, the $n$-point correlation function is
invariant under rigid-body rotation of the spatial coordinates.
For $n \le d$, this implies that $S^{(i)}_n$ depends only on the
distances $x_{ij}=|{\bf x}_{ij}|$ ($1 \le i < j \le n$).

In general, an infinite set of $S_n$ with $n=1, 2, 3 ...$ is
required to completely determine the microstructure and thus, the
effective properties of a heterogeneous material \cite{SalBook}.
Specifically, the effective property of interest can written as an
infinite series involving integrals of such correlation functions.
In practice, a complete knowledge of all of the
$S_n$ is never available. However, it has been shown that certain
approximations that can be regarded as resummations of the infinite series expansion
that incorporate lower-order $S_n$ (e.g., $S_2$, $S_3$ and $S_4$) can
provide accurate estimates of the effective properties
\cite{SalBook}. We note that since only certain weighted integrals
of the correlation functions are needed, excellent approximations
of effective properties can be obtained even without computing all of the
$S_n$ explicitly. We will discuss these approximations in detail
in Sec. 3.1.


\subsection{Pore-Size Probability Density Function}


The pore-size probability density function $P(\delta)$ first
rose to characterize the pore space in porous media
\cite{prager63} and was then generalized to characterize any
heterogeneous material \cite{SalBook}. For a statistically
homogeneous and isotropic material, $P(\delta)d\delta$ gives the
probability that a randomly chosen point in the pore space lies at
a distance between $\delta$ and $\delta+d\delta$ from the nearest
point on the pore-solid interface. Since it is a probability
density function with dimensions of inverse length, we have
$P(\delta) \ge 0$ for all $\delta$, and it normalizes to unity,
i.e.,
\begin{equation}
\label{eq_Pnormal}
\int_0^{\infty}{P(\delta)d\delta} = 1.
\end{equation}
At extreme values of $P(\delta)$, we have that
\begin{equation}
\label{eq_Pextreme}
P(0) = s/\phi_1, \quad P(\infty) = 0,
\end{equation}
where $s$ is the pore-solid interface area per unit volume and
$\phi_1$ is the volume fraction of the pore space. Therefore,
$s/\phi_1$ is the interface area per unit pore volume. The moments
of $P(\delta)$, defined as
\begin{equation}
\label{eq_Pmoments}
<\delta^n> = \int_0^{\infty}\delta^n P(\delta)d\delta,
\end{equation}
provide useful characteristic length scales of the pore space in
the material. Certain lower-order moments of $P(\delta)$ also
arise in bounds on the mean survival time $\tau$ \cite{prager63,
ave91}, which we will discuss in Sec. 3.


It is very difficult to obtain analytical expression of
$P(\delta)$ for a general polymer network. For networks composed
of very long and stiff polymer fibers, Ogston \cite{ogston58}
derived an expression for $P(\delta)$, i.e.,
\begin{equation}
\label{eq_POgston}
P(\delta) = \frac{2\phi_2 (\delta+a)}{a^2}e^{-\phi_2 (\delta+a)^2/a^2},
\end{equation}
where $\phi_2 = 1-\phi_1$ is the volume fraction of the fibers and
$a$ is the fiber radius. For $\delta = 0$, Eq.~(\ref{eq_POgston})
gives
\begin{equation}
\label{eq_P0}
P(0) = s.
\end{equation}
Comparing Eq.~(\ref{eq_P0}) and Eq.~(\ref{eq_Pextreme}), it is
clear that the Ogston expression for $P(\delta)$ can only provide
good estimates of it at very low fiber volume fractions, i.e.,
$\phi_2 \rightarrow 0$ and $\phi_1 = 1 -\phi_2 \approx 1$. In
general, the Ogston expression will underestimate $P(\delta)$ at
intermediate $\delta$, as we will show in Sec. 4.


Given any network microstructure, the associated $P(\delta)$ can
be numerically computed. Specifically, one generates many test points
that are randomly distributed in the pore space exterior to the
fibers and compute the distances from each test point to the
nearest fiber surface. This amounts to finding the largest test
sphere centered at the test point that is entirely in the pore
space. The resulting distances are binned to obtain a probability density
function, which is then normalized to give $P(\delta)$.

\section{Mean Survival Time and Effective Diffusion Coefficient}

\subsection{Theoretical Techniques}

\subsubsection{Mean Survival Time}

Consider the steady-state problem of diffusion of macromolecules
which are absorbed upon contacting the network fibers. This
implies that the rate of production of the macromolecules per unit
volume $G$ is exactly compensated by the rate of removal by the
traps. Locally, the process is described by the following Poisson
equation \cite{SalBook}
\begin{equation}
\label{eq_trap_Poi}
D_1 \Delta c = - G ~{\mbox {in ${\cal V}_1$}}, \quad c = 0 ~{\mbox{ on $\partial {\cal V}$}},
\end{equation}
where $c$ is concentration of the macromolecule, $D_1$ is
diffusion coefficient of the macromolecule in the pore space ${\cal V}_1$
and $\partial {\cal V}$ is the pore-fiber interface. 
The boundary condition that $c=0$ on $\partial {\cal V}$ assumes 
a perfectly absorbing interface. i.e., a diffusion-controlled reaction. 
This boundary condition can easily be relaxed to take into account 
partially absorbing interfaces \cite{SalBook, ave91}.

An important quantity is the mean survival time $\tau$ associated
with a macromolecule which is the average time that a diffusing
macromolecule spends in the pore space before it gets trapped by
the fibers. In many medical applications, the
efficiency of a drug strongly depends on the ability of the drug
macromolecules to diffuse through the extracellular space mainly
composed of collagen without getting trapped by the fibers
\cite{comper96}. It is noteowrthy that nuclear magnetic resonance (NMR)
relaxation in porous media, a widely used technique for biomedical
imaging, yields an NMR relaxation times from which one can extract 
the mean survival time we consider here \cite{SalBook}.

The mean survival time $\tau$ is inversely proportional to the
trapping constant $\gamma$, i.e.,
\begin{equation}
\label{eq_gamma}
\tau = \frac{1}{\gamma \phi_1 D_1},
\end{equation}
where $\gamma$ is defined via
\begin{equation}
G = \gamma D_1 <c>
\end{equation}
where $<c>$ is the ensemble-averaged concentration field.

The optimal lower bound on the mean survival time $\tau$ that
incorporates information on the pore space in terms of the first
moment of the pore-size probability density function is given by
\cite{ave91, prager63}
\begin{equation}
\label{eq_tau_lower_bd}
\tau \ge \frac{<\delta>^2}{D_1},
\end{equation}
where $<\delta> = \int_0^{\infty}\delta P(\delta)d\delta$. It can
be seen from Eq.~(\ref{eq_tau_lower_bd}) that $\tau D_1$ can
provide an estimate of the average pore size of the network. A
corresponding upper bound on the trapping constant can be obtain
by substituting Eq.~(\ref{eq_gamma}) into
Eq.~(\ref{eq_tau_lower_bd}).

\subsubsection{Effective Diffusion Coefficient}


Consider macromolecules diffusing between the fibers that are not absorbed by the fibers. The
Brownian motions of the macromolecules are hindered by the fibers in the network which
results in an effective diffusion coefficient $D_e$ smaller than
that of the pure solvent in the pore space $D_1$. By mathematical
analogy, the problem of macromolecular diffusion in the pore space
exterior to the fibers is equivalent to the electrical or thermal
conduction problem in the pore space with perfectly
insulating fibers. Specifically, the local ``flux'' ${\bf J}({\bf
x})$ of macromolecules is proportional to a local ``intensity''
${\bf E}({\bf x})$ which is the negative gradient of the
macromolecule concentration field $c({\bf x})$, i.e.,
\begin{equation}
\label{eq_flux}
{\bf J}({\bf x}) = {\bf D}({\bf x})\cdot {\bf E}({\bf x}) = -{\bf D}({\bf x})\cdot \nabla c({\bf x}),
\end{equation}
where
\begin{equation}
{\bf D}({\bf x}) = \left \{{\begin{array}{c} D_1 {\bf I}, \quad {\bf x}\in {\cal V}_1 \\ 0, \quad {\mbox {otherwise}}\end{array}}\right.
\end{equation}
and ${\bf I}$ is the unit second-order tensor. Under steady-state
conditions with no source and sink terms, the conservation of
macromolecules requires that ${\bf J}({\bf x})$ be solenoidal \cite{SalBook},
i.e.,
\begin{equation}
\nabla \cdot {\bf J}({\bf x}) = 0.
\end{equation}


If ${\bf D}({\bf x})$ in Eq.~({\ref{eq_flux}}) is replaced by the
local conductivity tensor ${\bf \sigma}({\bf x})$, one obtains the
local governing equations for conduction problems. We would like
to emphasize that although the diffusion problem and conduction
problem are equivalent in their mathematical formulations, there
is an important distinction between the effective diffusion
coefficient $D_e$ and the effective conductivity $\sigma_e$. For
the conduction problem, although the fiber phase is insulating, its
contribution to $\sigma_e$ is still explicitly considered. For
example, suppose that one randomly places test particles and
tracks their Brownian motions to compute $\sigma_e$. There is a
fraction of total number of particles $\phi_2$, which are
initially in the insulating fiber phase and will be trapped there
forever. Clearly these test particles, which have a zero diffusion
coefficient, are taken into account in the ensemble average for
$\sigma_e$. On the other hand, only the test particles in the pore
space are considered in order to compute $D_e$. Therefore, $D_e$ and
$\sigma_e$ for the same microstructure are related to one another
via the following relation:
\begin{equation}
\label{eq_D_Sigma}
\frac{D_e}{D_1} = \frac{1}{\phi_1} \frac{\sigma_e}{\sigma_1} =  \frac{1}{(1-\phi_2)} \frac{\sigma_e}{\sigma_1}.
\end{equation}

In the following, we present rigorous bounds and various analytical
approximations for the effective diffusion coefficient $D_e$. These
results were reported in literature for the effective conductivity
$\sigma_e$ of a general heterogeneous material. Here, we modify
them according to Eq.~(\ref{eq_D_Sigma}) to obtain expressions for
$D_e$.

\bigskip
\noindent{\it Hashin-Strikman Upper Bound}
\bigskip

For a two-phase heterogeneous material with an arbitrary but isotropic
microstructure in which one of the phases (e.g., phase 2) is
insulating, the Hashin-Strikman (HS) upper bound for the effective
diffusion coefficient $D_e$ is given by
\begin{equation}
\label{eq_HS_bd}
\frac{D_e}{D_1} = \frac{2}{2+\phi_2}.
\end{equation}
The HS lower bound in this case is trivially zero for all values
of $\phi_2$ \cite{SalBook}.

Although only volume fractions explicitly appear in the
expression, it has been shown that HS bound is the optimal bound
given the two-point information of the isotropic microstructure, i.e., $S_2$
\cite{SalBook}. Specifically, it is shown that the HS bounds are
realizable for a special class of ``coated sphere'' model
microstructures \cite{SalBook}. However, it is clear that such
two-point information is far from a complete characterization of
the network microstructure. Therefore, it can be expected that HS
upper bound is not tight, as we will show in Sec. 4.

\bigskip
\noindent{\it Low-Density Approximation}
\bigskip

For a heterogeneous material with microstructure composed of
well-defined inclusions (such as spheres, ellipsoids or cylinders)
in a matrix, the effective properties of the material can be
written as a power series of the volume fraction $\phi_2$ of the
inclusions \cite{SalBook}. When $\phi_2$ is sufficiently small,
i.e., in the low-density limit, truncating the power series
through the first-order in $\phi_2$ can provide a reasonable
estimate of the effective properties of interest.

For fiber networks, we consider that each fiber is an elongated
prolate spheroid in the ``needle'' limit. In such cases, the effective
diffusion tensor for dilute suspensions of orientated needles is
given by
\begin{equation}
{\bf D}_e = \left [{\begin{array}{c@{\hspace{0.3cm}}c@{\hspace{0.3cm}}c}
(D_e)_{11} & 0 & 0  \\
0 & (D_e)_{22} & 0 \\
0 & 0 & (D_2)_{33}\\ \end{array}} \right],
\end{equation}
where
\begin{equation}
\begin{array}{c}
(D_e)_{11} = (D_e)_{22} = D_1 [1 - \phi_2 + O(\phi_2^2)], \\
(D_e)_{33} = D_1 [1 + O(\phi_2^2)].
\end{array}
\end{equation}
For statistically isotropic materials such as suspensions of
randomly orientated needles in a matrix, the effective diffusion
coefficient is the average of the three principal components of
the tensor ${\bf D}_e$, i.e.,
\begin{equation}
\label{eq_De_Dilute}
\frac{D_e}{D_1} = 1 - \frac{2}{3} \phi_2 + O(\phi_2^2).
\end{equation}

Physically, Eq.~(\ref{eq_De_Dilute}) corresponds to the diffusion
of Brownian particles in a matrix with a single infinitely long
fiber, which clearly underestimates $D_e$ for actual biopolymer
networks. However, for certain microstructures such as periodic
arrays of parallel cylinders at low volume fractions,
Eq.~(\ref{eq_De_Dilute}), which we call the {\it low-density approximation}, 
can provide accurate estimates of
$D_e$, which can be used as benchmarks to test our
simulation results.

\bigskip
\noindent{\it Ogston Approximation}
\bigskip

An improved approximation for $D_e$ over the aforementioned
low-density approximation can be obtained if the contributions of
multiple fibers are taken into account simultaneously. This can be
done by using the idea of an ``influence cylinder'' associated
with each fiber introduced by Ogston \cite{ogston73}.
Specifically, consider a ``coated cylinder'' with outer radius $b$
and inner radius $a$ (i.e., the radius of the fiber). One can
easily compute the local effective diffusion coefficient $D_L(b)$
associated with the coated cylinder, i.e.,
\begin{equation}
\label{eq_D_ccylinder}
\frac{D_L(b)}{D_1} = \frac{1}{1+a^2/b^2} = \frac{1}{1+\phi_2(b)},
\end{equation}
where $\phi_2(b)$ is local volume fraction of the fiber in the coated cylinder.

Now consider that the global $D_e$ of a network is a weighted average
of the local $D_L(b)$ for the influence cylinders associated with
each fiber, which leads to the relation
\begin{equation}
\label{eq_D_cglobal}
\frac{D_e}{D_1} = \int_a^{\infty}f(b)\frac{D_L(b)}{D_1}db,
\end{equation}
where $f(b)$ is the influence cylinder distribution function
\cite{joha91c}. Ogston and coworkers assume that the influence cylinders contribute to
the global $D_e$ in the same way they contribute to the pore-size
probability density function $P(\delta)$, i.e.,
\begin{equation}
\label{eq_P_ccylinder}
P(\delta) = \int_a^{\infty}f(b)g(b, \delta)db,
\end{equation}
where $g(b, \delta)$ is the local pore-size distribution associated
with a coated cylinder with outer radius $b$ given by
\begin{equation}
g(b, \delta) = \frac{2(a+\delta)}{b^2}H[\delta-(b-a)]
\end{equation}
and $H(x)$ is the Heaviside step function, equal to unity for
$x > 0$ and zero otherwise. Then $f(b)$ can be obtained
by de-convolution of Eq.~(\ref{eq_P_ccylinder}) with a knowledge of
$P(\delta)$, either from direct numerical sampling or theoretical considerations.


\bigskip
\noindent{\it Torquato Approximation}
\bigskip

Torquato has derived an expansion of the effective conductivity
$\sigma_e$ of any two-phase heterogeneous materials in terms of
the contrast (difference) between the conductivities of the
individual phases \cite{Sal85}. He showed that the [2,2]
Pad{\'{e}} approximant of the so-called ``strong-contrast''
expansion, which incorporates up to 4-point microstructural
information involving integrals over $S_2$, $S_3$ and $S_4$ can
provide excellent estimates of $\sigma_e$ for a wide range of
model microstructures \cite{Sal85}.

Here, we present the modified 4-point Pad{\'{e}} approximation of
the effective diffusion coefficient $D_e$ for collagen networks,
which is henceforth referred to as the Torquato approximation,
i.e.,
\begin{equation}
\label{eq_TApprox}
\frac{D_e}{D_1} = \frac{1}{1-\phi_2}\frac{(1+\frac{1}{2}\frac{\gamma_2}{\zeta_2}-\frac{1}{2}\zeta_2)
+(-1+\frac{1}{2}\zeta_2-\frac{1}{2}\frac{\gamma_2}{\zeta_2})\phi_2}
{(1+\frac{1}{2}\frac{\gamma_2}{\zeta_2}-\frac{1}{2}\zeta_2)
+(\frac{1}{2}+\frac{1}{2}\zeta_2+\frac{1}{4}\frac{\gamma_2}{\zeta_2})\phi_2},
\end{equation}
where the parameter $\zeta_2$ is a weighted integral that involves
correlation functions $S_1$, $S_2$ and $S_3$ of the fiber phase; and
the parameter $\gamma_2$ is a weighted integral that involves
correlation functions $S_1$, $S_2$, $S_3$ and $S_4$ of the fiber phase.
The readers are referred to Ref.~\cite{Sal85} for detailed discussions
of these parameters. Useful rigorous inequalities relating $\zeta_2$ and $\gamma_2$
for three-dimensional microstructure are the following \cite{Sal85}:
\begin{equation}
\label{eq_parambd}
-1 \le \gamma_2/\zeta_2 \le 1-2\zeta_2.
\end{equation}

It is in general nontrivial to compute the 3-point and 4-point
correlation functions $S_3$ and $S_4$, even for relatively simple
model microstructures (e.g., dispersions of spheres), to obtain
exact values of $\zeta_2$ and $\gamma_2$. However, has been shown
that simplfied estimates of these parameters based on limited
microstructural information in conjunction with
Eq.~(\ref{eq_TApprox}) can lead to excellent approximations for
the effective properties \cite{Sal85, SalBook}. Here, we consider
the low-density approximation of $\zeta_2$ for a prolate spheroid
in the needle limit and only keep the leading order term, i.e.,
$\zeta_2 = 1/4$ \cite{SalBook}. The inequalities given in
Eq.~(\ref{eq_parambd}) then become
\begin{equation}
\label{eq_parambd2}
-1 \le \gamma_2/\zeta_2 \le 1/2.
\end{equation}
It has been shown that for dispersions of spheres, $\gamma_2 = 0$
can provide a very accurate approximation formula for the
effective conductivity of a wide range of sphere volume fractions
\cite{Sal85}. We will show in Sec. 4 that a proper choice of
$\gamma_2$ value that is near its lower bound can provide an
excellent approximation for the effective diffusion coefficient
associated with biopolymer networks.


\subsection{First-Passage-Time Simulation Techniques}


A straightforward way of numerically studying Brownian motion is
to simulate the exact zig-zag path of a diffusing particle (e.g.,
see Ref.~\cite{joha93b}). However, it is clear that this direct
approach is not efficient means of obtaining effective diffusive
properties, since the details of the diffusion paths are averaged
out and do not contribute to the effective behavior. Moreover,
one needs to consider a wide range of step sizes associated with
each random Brownian jump to extrapolate the results to the case
of infinitesimal small step size.



An alternative but much more computationally efficient approach is
the first-passage-time (FPT) simulation technique introduced by
Torquato and coworkers \cite{Sal89APL, kimall, kimall2, kimall3,
kimJCP}. The key idea of the FPT approach is not to simulate the
details of the zig-zag diffusion paths but rather consider the
average time that it takes a Brownian particle to ``jump''
directly  to a random location on the surface of the largest
imaginary sphere that is centered at the original position of the
particle and entirely within the solvent (i.e., the pore phase).
The imaginary sphere is referred to the ``first-passage sphere''
(FPS), whose radius is $R$. It can be shown that the mean time
$\tau_{\mbox{\tiny FPS}}$ for a Brownian particle, which is
initially at the center of the FPS and takes a complicated zig-zag
path to hit the surface of the FPS is, in three dimensions, given
by
\begin{equation}
\label{eq_t_fps} \tau_{\mbox{\tiny FPS}} = R^2/(2dD_1),
\end{equation}
where $d=3$ is the space dimension.


When the particle is very close to the fiber surface, i.e., the
distance $r$ from the particle centroid to the fiber surface is smaller than
a prescribed tolerance $\Delta$, we consider that the particle
hits the fiber surface and is reflected back. The FPS in this case
encloses both the pore phase and fiber phase.
Suppose that the FPS centered at the Brownian particle centroid possess a radius $R$,
the associated time $\tau_{\mbox{\tiny REF}}$ that the Brownian particle takes
to hit the fiber surface, be reflected back and hit the FPS can be
estimated by
\begin{equation}
\label{eq_t_ref}
\tau_{\mbox{\tiny REF}}(r) = \frac{R^2}{6 D_1} \frac{V_1+V_2}{V_1}\left [{ 1 +\frac{1}{2} \left ({\frac{r}{R}}\right)^2 \\
-\frac{1}{2}\sum_{m=0}^{\infty}C_{2m+1}\left ({\frac{r}{R}}\right)^{2m+1}}\right ],
\end{equation}
where $0 \le r \le R$, $V_1$ and $V_2$ are the volume of the pore phase
and the volume of the fiber phase enclosed in the FPS, respectively, and
\begin{equation}
C_{2m+1} = \frac{(-1)^{m+1}(2m)!}{2^{2m+1}(m!)^2}\frac{3(4m+3)}{(2m-1)(m+2)(m+1)}.
\end{equation}
Equation (\ref{eq_t_ref}) was first derived by Torquato and
coworkers \cite{kimall, kimall2, kimall3}, and it has been shown
to provide excellent approximation of the exact $\tau_{\mbox{\tiny
REF}}$ for any local geometry when $r \ll R$ and $R$ is smaller
than the diameter of the fiber. The readers are referred to
Ref.~\cite{kimall} and the references therein for additional
details.

To compute $\tau_{\mbox{\tiny REF}}$ using Eq.~(\ref{eq_t_ref}), a
key step is to evaluate the intersection volume between a sphere
and a cylinder (i.e., $V_2$), the details of which are given in
the Appendix. In the rare case that the particle is close to a
cross link (junction of several fibers), $V_2$ is the volume of
the fiber junction enclosed in the FPS, which is computed by Monte
Carlo sampling \cite{kimall3}. For example, one randomly places
test points in the FPS and computes the fraction of times that the
point falls into the vicinity of the fiber junction.

To obtain $D_e$, one considers an ensemble of Brownian trajectories in the pore
space. When a diffusing particle is sufficiently far away from the fiber surface, one
constructs the largest FPS of radius $R$ around the diffusing particle which
just touches the fiber surface. The particle then jumps in one step to a
random point on the surface of the FPS and the process is repeated, each time
keeping track of $R^2_i$, until the particle is within a prescribed very small
distance $\Delta$ to the fiber surface (see Fig. 3). At this point in time, the particle
is considered to hit the fiber and then is reflected back. Thus, one keeps track of
the radius $R_j$ of the FPS that encloses both the fiber phase and the pore phase
and computes the associated time $\tau_{\mbox{\tiny REF}}(R_j)$. The expression for
effective diffusion coefficient $D_e$ is then given by
\begin{equation}
\label{eq_De_FPT}
\frac{D_e}{D_1} =\left \langle{ \frac{\sum_i R_i + \sum_j R_j}{\sum_i R_i + 6D_1 \sum_j \tau_{\mbox{\tiny REF}}(R_j)} }\right \rangle,
\end{equation}
where $\tau_{\mbox{\tiny REF}}(R)$ is given by Eq.~(\ref{eq_t_ref}) and $<.>$
denotes the ensemble average over many Brownian particles. In our
simulations, we use $N=5~000$ Brownian particles.

The mean survival time $\tau$ can be obtained in a similar way \cite{Sal89APL}. Specifically,
one constructs the first-passage-time path composed of many jumps to the surface of FPS
associated with a Brownian particle and keeps track of $R_i$ for each FPS.
When the particle is within $\Delta$ to the fiber surface, it is considered trapped
by the fiber. Thus, the mean survival time can be computed via
\begin{equation}
\label{eq_tau_FPT}
\tau = \left \langle {\sum_i R_i/D_1} \right \rangle,
\end{equation}
where $<.>$ denotes ensemble average over many Brownian particle. In our
simulations, we use $N=5~000$ Brownian particles.

We note that in the aforementioned first-passage-time simulation
technique, we consider that the Brownian particle is ``point'' particle
with zero diameter. For finite-sized particles diameter $d_{\mbox{\tiny P}}$, it has been shown
that one can still consider ``point'' particles in a network microstructure
with the diameter of the fibers $d_{\mbox{\tiny F}}$ dilated by $d_{\mbox{\tiny P}}$ \cite{kimJCP, SalJCP91}.

\section{Results}

Our data are ``graph representations'' of collagen type I networks
\cite{lind10} with final collagen concentrations of 1.0 mg/ml, 2.0
mg/ml and 4.0 mg/ml. The fibers roughly possess a
circular cross section of diameter $d_{\mbox{\tiny F}} = 1.0\times
10^{-7}$ m \cite{lind10}. The average fiber lengths
$\ell_{\mbox{\tiny F}}$ for the networks with the three collagen
concentrations are respectively $1.96 \times 10^{-6}$m, $1.81
\times 10^{-6}$m, and $1.28 \times 10^{-6}$m. The corresponding
volume fractions of the fibers are respectively $1.7\times
10^{-3}$, $2.4\times 10^{-3}$, and $5.2\times 10^{-3}$. The
tolerance $\Delta$ described in Sec. 3 is chosen to be $\Delta =
5\times 10^{-3}d_{\mbox{\tiny F}} = 5.0\times 10^{-10}$ m. Since
we do not consider hydrodynamic effects in our simulations, we
only consider particles with diameter $d_{\mbox{\tiny P}}$
comparable to the fiber diameter $d_{\mbox{\tiny F}}$, i.e.,
$d_{\mbox{\tiny P}} \le d_{\mbox{\tiny F}}$. For large
$d_{\mbox{\tiny P}}$, it has been shown that the hydrodynamic
effects on Brownian motion are significant \cite{styl10}. The
results reported below are ensemble averages of three independent
network configurations at each collagen concentration.

\subsection{Pore-Size Probability Density Function}


The pore-size probability density functions $P(\delta)$ for the collagen
network at three concentrations are numerically computed as
described in Sec. 2. The obtained $P(\delta)$ are shown in
Fig.~4 and compared to the corresponding Ogston expressions
[Eq.~(\ref{eq_POgston})] at the same fiber volume fractions.

It can be clearly seen that the Ogston expression of $P(\delta)$
overestimates the number of intermediate pores and underestimates
the number of large pores in the system. This is because in the
derivation of Eq.~(\ref{eq_POgston}), it is assumed that the
network is composed of fibers with very long persistence length.
For the collagen networks we study, the average fiber lengths
$\ell$ are less than twice of the corresponding averaged pore size
$<\delta>$ [defined in Eq.~(\ref{eq_Pmoments})], which are
respectively $1.22 \times 10^{-6}$m, $0.998 \times 10^{-6}$m, and
$0.684 \times 10^{-6}$m for collagen concentrations 1.0 mg/ml, 2.0
mg/ml and 4.0 mg/ml. Therefore, the long-fiber-length assumption
for the Ogston expression is not true here.

The $P(\delta)$ data will be employed to compute the lower bound
on the mean survival time $\tau$ [Eq.~(\ref{eq_tau_lower_bd})] and
to compute the Ogston approximation for the effective diffusion
coefficient $D_e$ [Eq.~(\ref{eq_D_cglobal})] in the following
sections.

\subsection{Mean Survival Time}



The mean survival time $\tau$ is computed using the
first-passage-time technique described in Sec. 2. Figure 5
shows the scaled dimensionless mean survival time $\tau
D_1/\ell_F^2$ for the collagen networks with three different
concentrations as a function of Brownian particle diameter. It can
be seen that as the collagen concentration increases, larger
particles are more easily get trapped by the fibers. This fact is
of great importance in cancer chemotherapy, which we will discuss
in Sec. 6.


As indicated in Sec. 3, the diffusion of finite-sized particles in
the original network is equivalent to the diffusion of point particles
in a properly dilated network, which possesses a higher fiber volume
fraction. Figure 6 shows the scaled mean survival time $\tau D_1/\ell_F^2$ for
the collagen networks with three different concentrations
as a function of the particle diameter. The pore-size lower bounds are
also shown. It can be seen that although the bounds are not sharp,
they do not deviate very much from the actual mean survival times.
We note that these bounds only incorporate partial information about the
pore-size probability density function $P(\delta)$, namely, the first moment
$<\delta>$ of $P(\delta)$. Therefore, one would expect that incorporating
the full information content of $P(\delta)$ would lead to good predictions of the effective
diffusive properties considered here. Indeed, we will show in the following section that the
a generalization of the Ogston approximation that employs the complete microstructural information
contained in $P(\delta)$ provides a very good estimate of $D_e$.

In Ref.~\cite{Yeong97}, Torquato and Yeong found a universal curve
for a scaled mean survival time $\tau$ for a wide range of
microstructures with different porosities, including various
random and ordered distributions of spheres and certain continuous
models. Specifically, the universal curve has the following form:
\begin{equation}
\label{eq_t_universal}
\frac{\tau}{\tau_0} = a_1 x + a_2 x^2,
\end{equation}
where $a_1$ and $a_2$ are constants and
\begin{equation}
\tau_0 = \frac{3\phi_2}{D_1 \phi_1 s^2}, \quad x = \frac{<\delta>^2}{\tau_0 D_1},
\end{equation}
and $s$ is the specific surface, i.e., the solid-pore interface
area per unit volume. For the class of microstructures they
studied, Torquato and Yeong found that $a_1 = 8/5$ and $a_2 =
8/7$.


For the collagen networks studied here, we find that
Eq.~(\ref{eq_t_universal}) also holds (see Fig. 7). However, the
constants are different from those obtained by Torquato and Yeong,
i.e, we find that $a_1 = 0.121$ and $a_2 = 1.88$ for the networks with
different concentrations. A possible reason for the difference in
the constants is that collagen networks do not belong to the
same class of microstructures studied in Ref.~\cite{Yeong97}, which, for example, do not contain
filamentary-like structures, as in the case of collagen fibers.
This implies that there could exist a more general scaling curve for
the mean survival time that incorporates both the networks and the
microstructures studied in Ref.~\cite{Yeong97}. Nonetheless, our
results enable one to efficiently estimate the properties of
collagen networks. In particular, given any of the three
quantities among the four quantities $\tau$, $\phi_1$, $s$ and
$<\delta>$, the remaining one can be estimated employing
Eq.~(\ref{eq_t_universal}).

\subsection{Effective Diffusion Coefficient}


The effective diffusion coefficient $D_e$ for various network
microstructures are computed using both the theoretical techniques
described in Sec. 3.1 and the first-passage-time technique
described in Sec. 3.2. Figure 8 shows $D_e$ for the fiber
networks with different collagen concentrations as a function of
the Brownian particle diameter. Similar to the case of the mean
survival time, as the collagen concentration $\phi$ increases, it becomes
more and more difficult for larger particles to diffuse in the
collagen.

Figure 9 shows $D_e$ as a function of the fiber volume fraction.
In addition to the results for the collagen networks, we also show
$D_e$ for a model microstructure composed of parallel cylinders
arranged on a square lattice. The Hashin-Strikman (HS) upper bound
and various approximations of $D_e$ discussed in Sec. 3.1 are
also shown in Figure 9. As we indicated earlier, since the HS
bound only incorporates the limited two-point information $S_2$,
it can not provide a good estimate of $D_e$. This is also evident
from the fact that the HS bound is realized by certain class of
``coated sphere'' model microstructures, which are clearly
topologically distinct from the network microstructures because
one of the phases is topologically disconnected. It can be seen
that the low-density approximation also underestimates $D_e$ for
the collagen networks. This is because it only considers the
effect of a single long fiber to the diffusing particles. In the
actual networks, the average fiber length is less than twice the
average pore size as we indicated earlier. Moreover, the
cross-links also significantly hinder the diffusion of the
particles. However, for the parallel-cylinder model at low volume
fractions, the low-density approximation should provide accurate
estimates, since in such cases, the diffusion of the particles is
only hindered by well separated single cylinders. Indeed, we find
that the approximation agrees very well with our simulation data,
which also verifies the accuracy of our simulations.

The Ogston approximation that incorporates the pore-size
information of the networks provides a much better estimate of
$D_e$ compared to the HS bound and the dilute approximation.
However, it still slightly underestimates $D_e$ for large
particles in networks at high collagen concentrations. This is
because the ``influence cylinders'' (see Sec. 3.1) are
associated with individual long fibers and the effects of
finite-fiber length and the cross-links are still not fully
incorporated. On the other hand, one can see that the Torquato
approximation agrees extremely well with the simulation data for
all volume fractions that we considered. Specifically, in
employing Eq.~(\ref{eq_TApprox}) we have chosen the 4-point
parameter value such that $\gamma_2/\zeta_2 = -0.925$. Note that
this value is very close to the lower bound value -1, which is not
very surprising the networks can be considered as a kind of
``limit'' microstructure. This does not mean the actual value of
$\gamma_2/\zeta_2$ if computed with a full knowledge of the
associated 3-point function $S_3$ and 4-point function $S_4$,
since the value of $\zeta_2$ we used is also an approximation. The
success of the Torquato approximation is due to the fact that
higher-order microstructural information that
possibly reflects the effects of the cross-links is already taken
into account by the 4-point parameter $\gamma_2$ in the expression
Eq.~(\ref{eq_TApprox}).


\section{Estimating Elastic Properties of Collagen Network Using Cross-Property Relations}

Since effective properties of heterogeneous materials reflect
certain microstructural information about the material, it is
possible to extract rigorously information about one physical
property given an accurate determination of a different effective
property obtained either experimentally or theoretically. Such
interrelationships are called {\it cross-property} relations
\cite{SalBook, cross3, cross4, cross5, cross1, cross2}. Rigorous cross-property
relations become especially useful if one property is more 
easily measured than another property.

In this section, we estimate the effective bulk modulus $K_e$
\cite{SalBook} of fluid-saturated collagen networks using the
effective diffusion coefficient $D_e$ computed here using the
cross-property relations. In particular, Gibiansky and Torquato
\cite{cross3, cross4, cross5} derived a nontrivial rigorous cross-property upper bound
$K_e^U$ on the effective bulk modulus $K_e$ of a fluid-saturated porous
material with an insulating solid phase given the effective
conductivity (equivalent to the effective diffusion coefficient)
of the material, i.e.,
\begin{equation}
\label{eq_cross} K_e \le K_e^U = K_{1e}-\frac{2\phi_1\phi_2^2
G_2(K_1-K_2)^2}{a[3a\phi_1 F-3a-2\phi_2 G_2]},
\end{equation}
where
\begin{equation}
\begin{array}{c}
a = \phi_2 K_1+\phi_1 K_2 + 4G_2/3, \\ K_{1e} = \phi_1 K_1 +
\phi_2 K_2 - \phi_1 \phi_2 (K_1-K_2)^2/a,
\end{array}
\end{equation}
and $\phi_1$ and $K_1$ are respectively the volume fraction and
bulk modulus of the fluid phase; $\phi_2$, $K_2$, $G_2$ are
respectively the volume fraction, bulk modulus and shear modulus
of the solid phase. Moreover, the ``formation factor'' $F$ in
Eq.~(\ref{eq_cross}) is given by
\begin{equation}
F = \frac{1}{\phi_1} \frac{D_1}{D_e}
\end{equation}
where $D_1$ is the diffusion coefficient of the fluid phase and
$D_e$ is the effective diffusion coefficient of the porous
material.

For the collagen networks considered here, the solid phase
corresponds to the collagen fibers. The bulk modulus $K_e$ of
fluid-saturated collagen networks at fiber volume fraction $\phi_2
= 0.005$ has been measured experimentally \cite{bulk1}, i.e., $K_e
\approx 2500$ Pa. The shear modulus of ``dry'' collagen networks
(i.e.., fiber networks without fluid) have been computed
numerically by Lindstr{\"{o}}m et al. \cite{lind10}, i.e., $G_2 =
24$ Pa. We use $K_2 = 10$ Pa for the ``dry'' network and $K_1 = 2$ GPa
for the fluid. Using the Torquato approximation for the effective
diffusion coefficient $D_e$, the upper bound value 
obtained from inequality (\ref{eq_cross}) is computed,
i.e., $K_e^U = 3530$ Pa, which provides a surprisingly good estimate of
$K_e$, given the fact that $K_e^U$ is a rigorous upper bound \cite{SalBook}.

\section{Conclusions and Discussion}

In this paper, we have quantitatively characterized the microstructure, 
the mean survival time $\tau$, and the effective
diffusion coefficient $D_e$ of collagen type I networks by
applying theoretical and computational techniques from the theory of  heterogeneous
materials. Specifically, we have computed the pore-size
probability density function $P(\delta)$ for the networks. We have
also employed a variety of theoretical approximation schemes 
for the effective conductivity of a two-phase
material to estimate the effective diffusion coefficient $D_e$ for
the networks. Such estimates include the low-density approximation,
the Ogston approximation, and the Torquato approximation, all of which
incorporate different levels of microstructural information about the
networks. The Hashin-Strikman upper bound on $D_e$ and the
pore-size lower bound on $\tau$ are used as benchmarks to test our
results. Moreover, we have generalize the efficient
first-passage-time techniques for Brownian-motion simulations in
suspensions of spheres to the case of network microstructures and
compute the associated $D_e$ and $\tau$. We have found a universal
curve for $\tau$ for the networks at different collagen
concentrations and have shown that the Torquato approximation
which takes into account higher-order microstructural information
can provide the most accurate estimate of $D_e$ for all collagen
concentrations among the employed approximation schemes. Our work
also demonstrates that employing the rich family of theoretical
and simulation techniques developed in materials sciences to
characterize biological systems (e.g., the heterogeneous host
microenvironment of tumors) suggested in Ref.~\cite{Sal11} is a
very promising approach worthy further exploration.

We have found that as the collagen concentration increases,
the diffusion of large particles in the collagen networks, and
thus the extracellular matrix (ECM) becomes more and more
difficult and it is easier for the diffusing particles to be
trapped by the fibers. This is a major problem associated with any
cancer chemotherapy, since drug macromolecules would get trapped
by collagen fibers without successfully diffusing to the target
site. It is known that a growing malignant tumor constantly
modifies the chemical composition of the collagen networks
composing its ECM \cite{tumor_ECM}. In addition, since a pressure
is built up as the tumor grows, the surrounding ECM is pushed and
compressed, leading to a higher collagen concentration in tumor
ECM than in normal tissues \cite{tumor_stress1, tumor_stress2,
tumor_model}. Therefore, it can be expected that the diffusion of
drugs to the tumors is really difficult. More efficient
chemotherapies trying to overcome these difficulties are being
developed \cite{tumor_chemo}.

We also applied a rigorous cross-property upper bound to
to estimate the effective bulk modulus $K_e$ of collagen networks from
a knowledge of the effective diffusion coefficient $D_e$ computed here.
The estimated value of $K_e$ agrees well with existing experimental data, 
given the fact that it is a rigorous upper bound.

In future work, we intend to generalize our simulation techniques
and theoretical approaches to investigate transport properties of
tissues with both collagen networks and various types of cells.
Specifically, we will focus on the effects of the cell shape and
the plasma membrane on the diffusion of macromolecules. In
addition, we will model the mechanical behavior of tissues using
the well-developed methods for heterogeneous materials. Progress
in these studies should deepen our understanding of the effects of
the host microenvironment on tumor growth and would lead to better
cancer treatment strategies.

In addition, we will apply cross-property relations to estimate
other physical properties of collagen networks from a knowledge of
the effective diffusive transport properties computed in this
paper. In particular, given the latter, we will
bound the fluid permeability \cite{cross1, cross2} for the
collagen networks studied here.

\ack
The authors are very grateful to S. B. Lindstr{\"{o}}m and D. Weitz
for providing the data of the collagen networks.
This project was supported by the National Center for Research Resources and the
National Cancer Institute of the National
Institutes of Health through Grant Number U54CA143803.
The content is solely the
responsibility of the authors and does not necessarily represent
the official views of the National Cancer Institute or the
National Institutes of Health.

\section*{Appendix: Intersection volume of a sphere with a cylinder}

Consider a sphere with radius $R_s$ centered at the surface of a cylinder
with radius $R_c$, the intersection volume $V_I$ of the sphere and the
cylinder for the case $R_c > R_s$ is given by \cite{ref_vol}
\begin{equation}
V_I = \frac{2}{3}\pi R_s^3 + \frac{4}{9\sqrt A}\left [{K(k)(A-B)(3B-2A)
+E(k)A(2A-4B) }\right],
\end{equation}
where $K(k)$ and $E(k)$ are elliptic integrals of the first and second kind, respectively, i.e.,
\begin{equation}
K(k) = \int_0^1 \frac{dz}{\sqrt{(1-z^2)(1-k^2 z^2)}}, \quad
E(k) = \int_0^1 dz \sqrt{\frac{1-k^2 z^2}{1-z^2}},
\end{equation}
and
\begin{equation}
A = 4R_c^2, \quad B = R_S^2, \quad k^2 = B/A.
\end{equation}

\newpage

\section*{Abbreviations list}
\begin{itemize}
\item ECM: extracelluar matrix

\item HS: Hashin-Strikman

\item NMR: nuclear magnetic resonance

\item FPS: first-passage sphere
\end{itemize}

\newpage

\section*{References}

\newpage

\section*{Figure Legends}
\begin{itemize}
\item Figure 1: Collagen networks. (a) Confocal microscope image of collagen-I at a final collagen
concentration 2.0mg/ml. The linear size of the image is approximately
150 $\mu$m. Image courtesy of S. B. Lindstr{\"{o}}m. (b) Three-dimensional
``graph'' representation of the collagen
network studied here. The linear size of the box is approximately 100 $\mu$m.

\item Figure 2: An illustration of sampling $P(\delta)$ of the collagen network in two dimensions.
Shown are the thin fibers that can possibly intersect as well as a
test point (red) and its associated largest sphere (blue) entirely within in the
pore space exterior to the fibers.

\item Figure 3: An illustration of the first-passage-time simulation technique in two dimensions.
Shown are the thin fibers that can possibly intersect and several first passage spheres.
Starting from an initial position, a diffusing particle jumps to a
random location on the surface of its associated first-passage sphere.
The process is repeated until the particle is very close to the
fiber surface. The the particle will hits the fiber and be reflected back to
the pore space.

\item Figure 4: The pore-size probability density function $P(\delta)$ for collagen networks
at different collagen concentrations.

\item Figure 5: The scaled dimensionless mean survival time $\tau D_1/\ell_F^2$ for the collagen
networks at different collagen concentrations as a function of the diffusing
particle diameter.

\item Figure 6: The scaled dimensionless mean survival time $\tau D_1/\ell_F^2$ for the collagen
networks at different collagen concentrations as a function of the diffusing
particle diameter and the associated pore-size lower bound.

\item Figure 7: Universal curve for the scaled mean survival time
$\tau/\tau_0$ versus $<\delta>^2/(\tau_0 D_1)$ for the collagen
networks at different collagen concentrations. It is seen that the
scaled mean survival time for different collagen networks 
collapse onto a single curve.

\item Figure 8: The dimensionless effective diffusion coefficient
$D_e/D_1$ for the collagen networks at different collagen
concentrations as a function of the diffusing particle diameter.

\item Figure 9: The dimensionless effective diffusion coefficient
$D_e/D_1$ for the collagen networks at different collagen
concentrations as a function of the fiber volume fraction. The
Hashin-Strikman upper bound and various analytical approximations
as discussed in Sec.~3.1 are shown and compared to the
simulation data.

\end{itemize}

\newpage

\section*{Glossary}
\begin{itemize}



\item \textit{Collagen}: a group of naturally occurring and the
most abundant proteins (biopolymers) found in animals, especially
in the flesh and connective tissues of mammals.

\item \textit{Heterogeneous material}: a material composed
different materials (e.g., a composite) or the same material in
different state (e.g., a polycrystal). The fluid-saturated
collagen networks studied here are special heterogeneous
materials.

\item \textit{Diffusion coefficient}: a proportionality constant
between the molar flux due to molecular diffusion and the gradient
in the concentration of the species or the driving force for
diffusion.

\item \textit{Mean survival time}: the average time that a
diffusing molecule spends in the fluid phase before it gets
trapped at the interface of the collagen fibers assuming a perfectly absorbing interface.

\item \textit{Bulk modulus}: a measure of a material's
resistance to uniform compression, defined as the ratio of the
infinitesimal pressure increase to the resulting relative decrease
of the material's volume.

\item \textit{Shear modulus}: a measure of a material's
resistance to shape deformation shear stress, defined as the ratio
of shear stress to the shear strain.

\item \textit{Correlation function}: a statistical descriptor of
the microstructure of a heterogeneous material, quantifying the
spatial correlations at different points in the microstructure.

\item \textit{First-passage-time}: the average time that it takes
for a diffusing particle to ``jump'' directly to a random location
on the surface of an imaginary sphere that is centered at the
original position of the particle and entirely within the solvent
region.

\item \textit{Cross-property relation}: an interrelationship that
enables one to extract rigorously information about one physical
property of a heterogeneous material given an accurate
determination of a different property.

\end{itemize}

\newpage

\begin{figure}[bthp]
\begin{center}
$\begin{array}{c@{\hspace{1.5cm}}c}\\
\includegraphics[height=4.25cm,keepaspectratio]{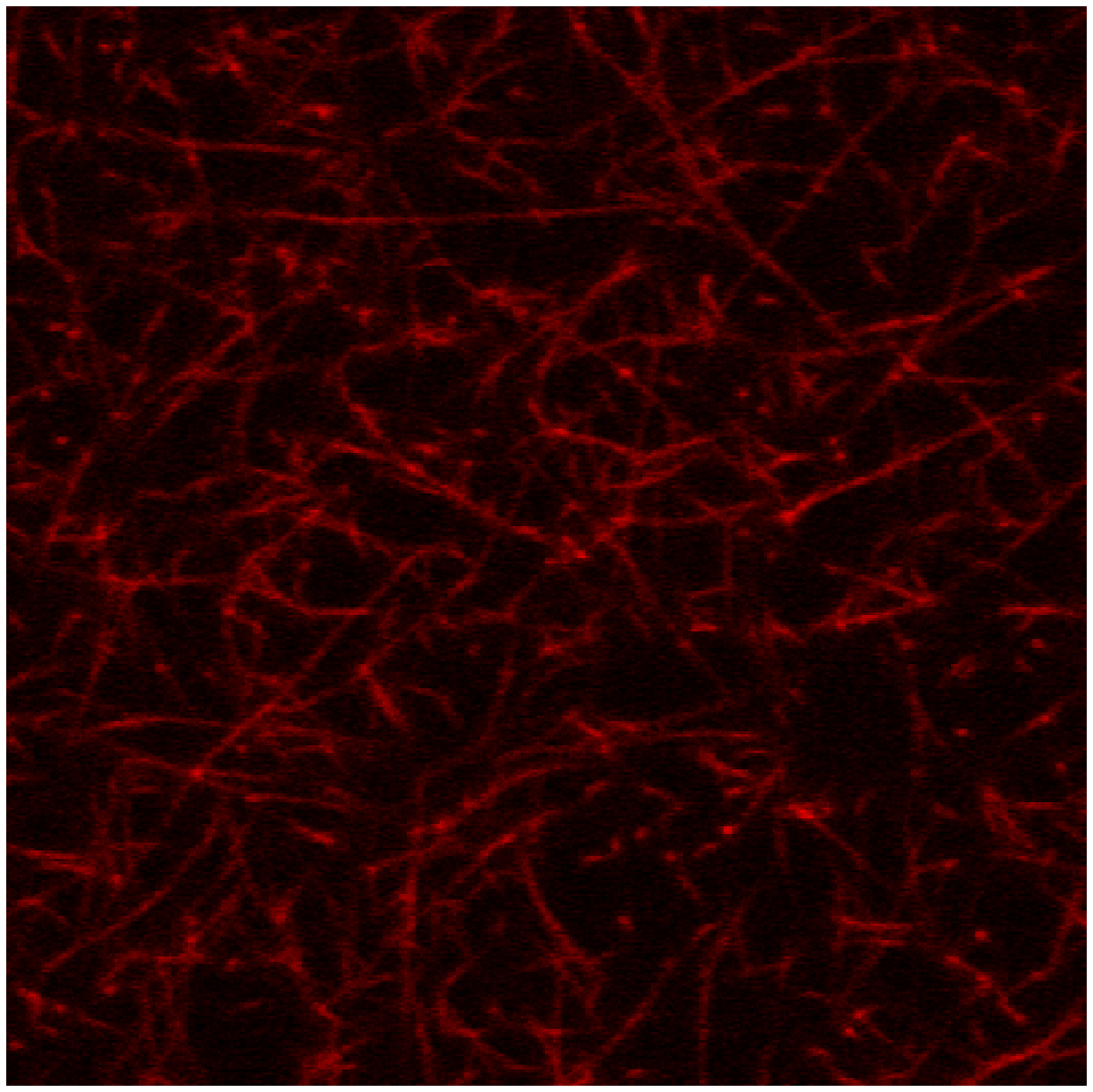} &
\includegraphics[height=5.25cm,keepaspectratio]{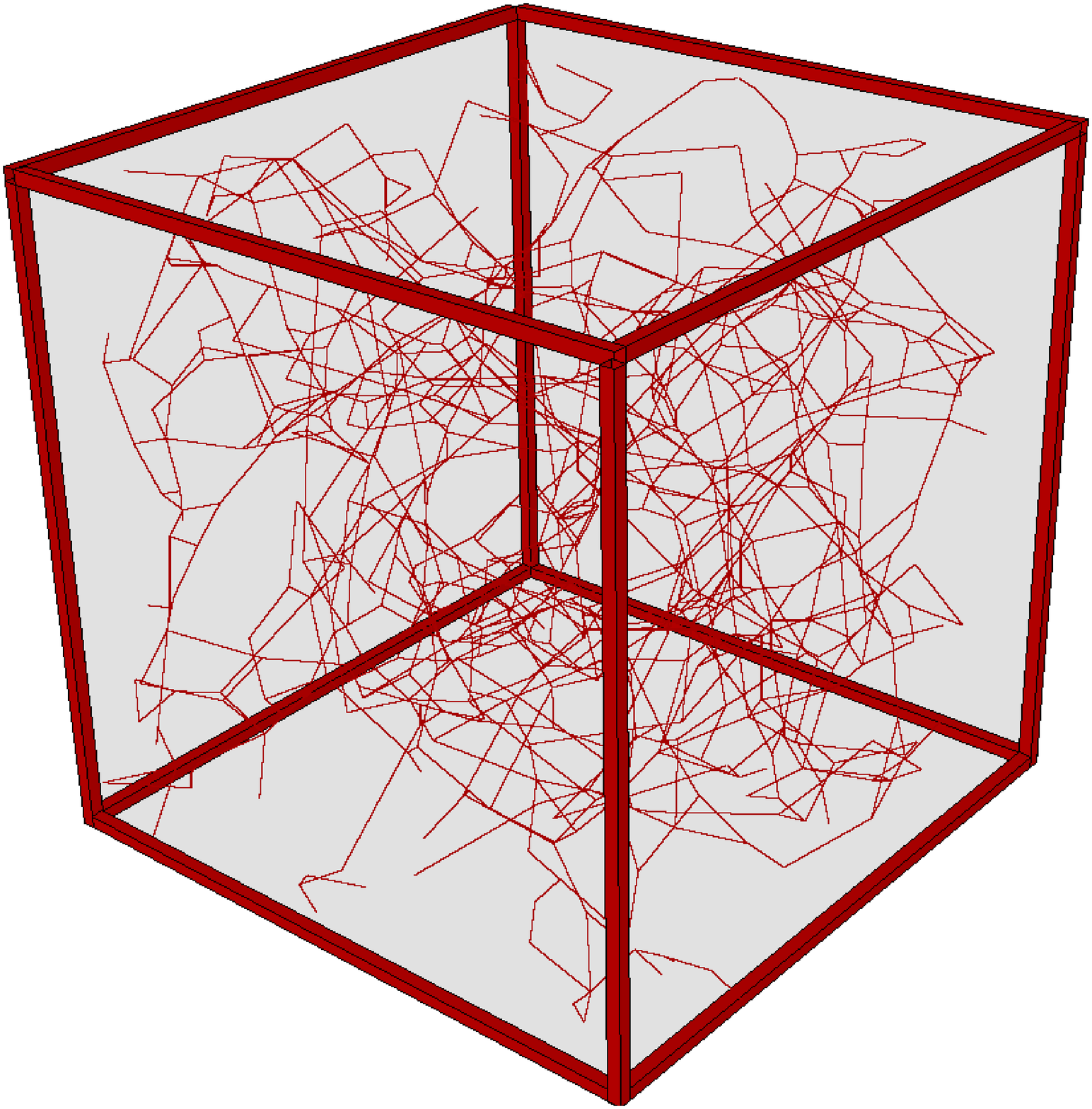} \\
\mbox{\bf (a)} & \mbox{\bf (b)}
\end{array}$
\caption{Jiao and Torquato}
\end{center}
\label{fig_coll_img}
\end{figure}
\bigskip
\bigskip
\bigskip
\bigskip

\begin{figure}[bthp]
\begin{center}
$\begin{array}{c}\\
\includegraphics[height=5.0cm,keepaspectratio]{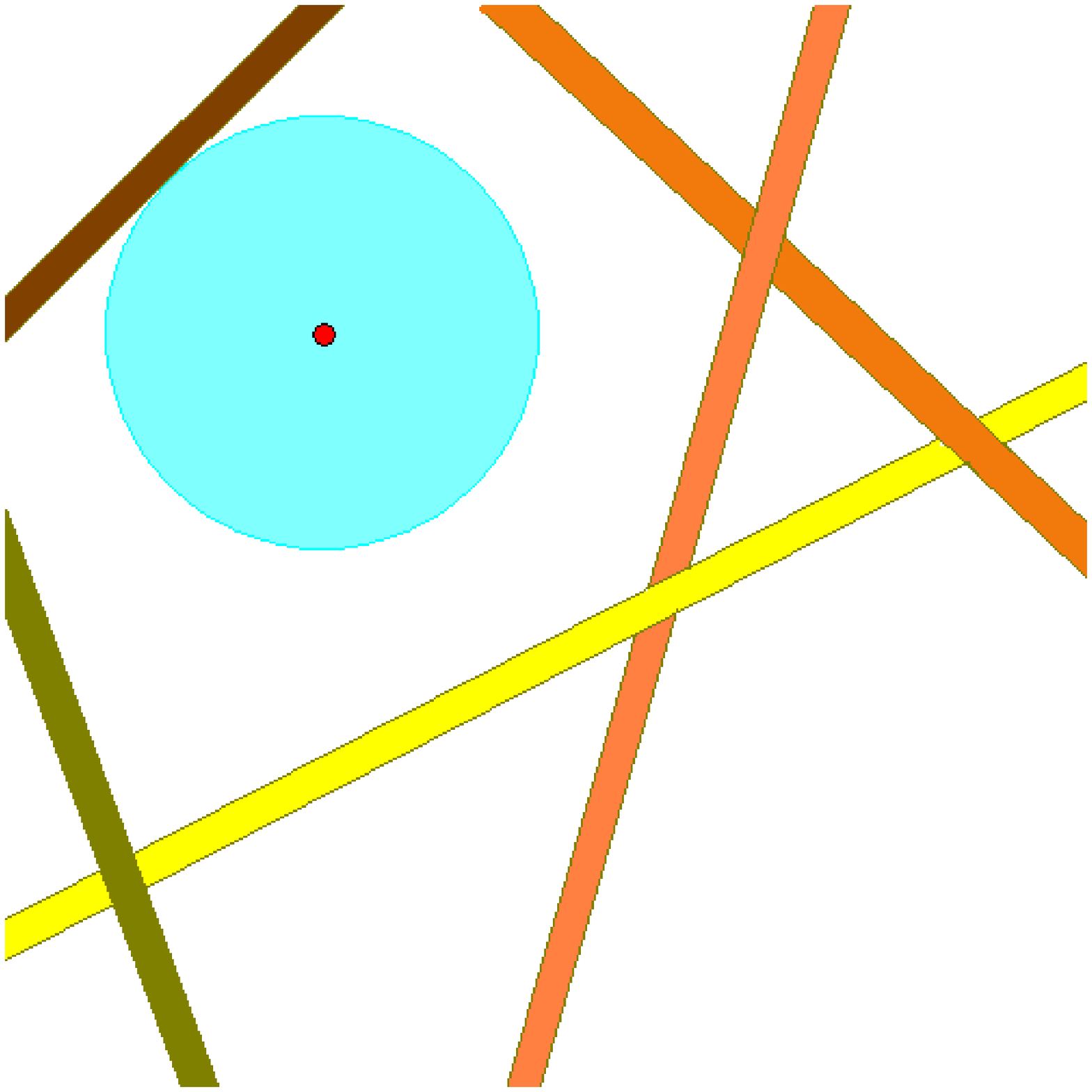} \\
\end{array}$
\caption{Jiao and Torquato}
\end{center}
\label{fig_Pz}
\end{figure}
\bigskip
\bigskip
\bigskip
\bigskip

\begin{figure}[bthp]
\begin{center}
$\begin{array}{c}\\
\includegraphics[height=5.0cm,keepaspectratio]{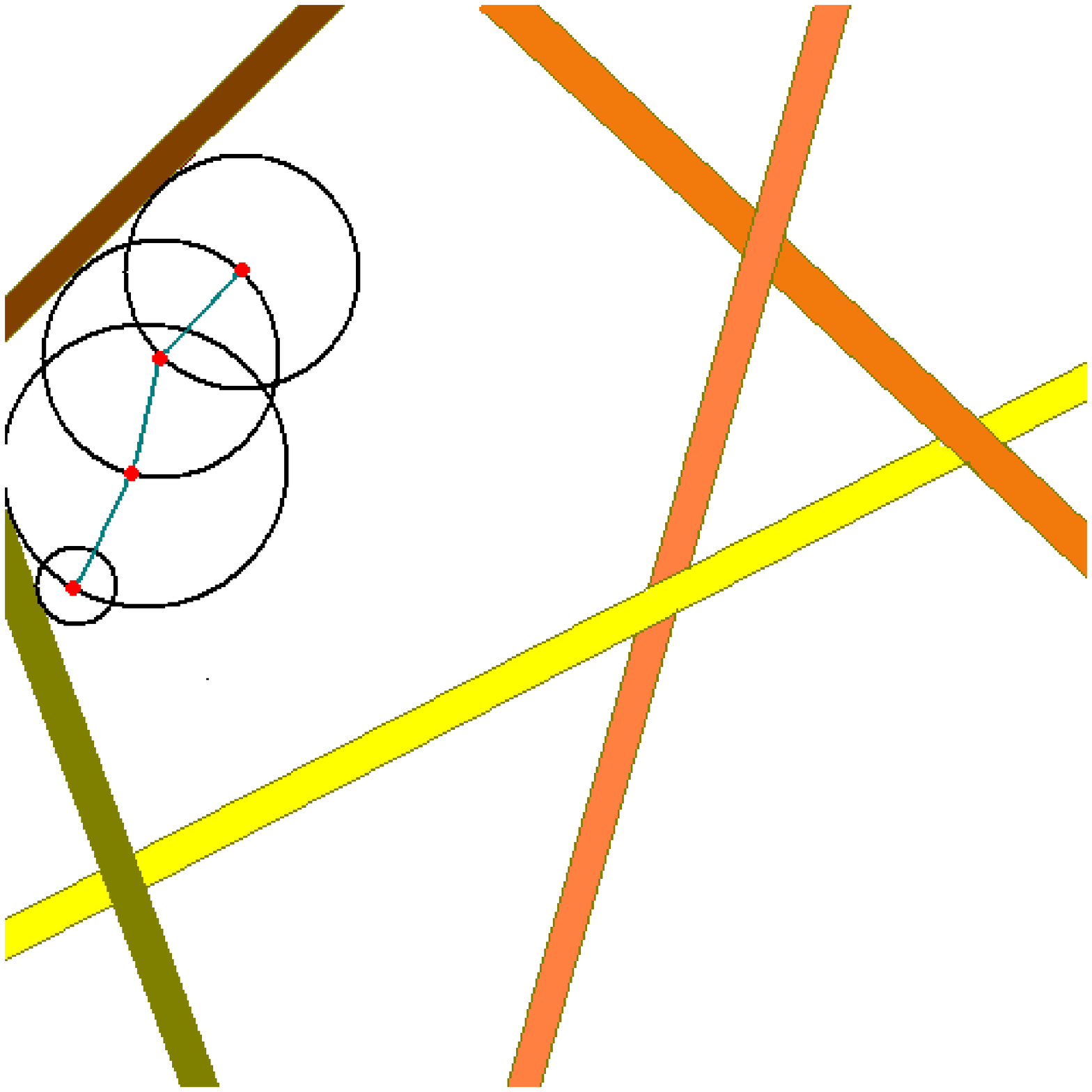} \\
\end{array}$
\end{center}
\caption{Jiao and Torquato}
\label{fig_FPT}
\end{figure}
\bigskip
\bigskip
\bigskip
\bigskip

\begin{figure}[bthp]
\begin{center}
$\begin{array}{c@{\hspace{0.6cm}}c@{\hspace{0.6cm}}c}\\
\includegraphics[height=4.5cm,keepaspectratio]{fig4a.eps} &
\includegraphics[height=4.5cm,keepaspectratio]{fig4b.eps} &
\includegraphics[height=4.5cm,keepaspectratio]{fig4c.eps} \\
\mbox{\bf (a)} & \mbox{\bf (b)} & \mbox{\bf (c)}
\end{array}$
\end{center}
\caption{Jiao and Torquato}
\label{fig_Pzdata}
\end{figure}
\bigskip
\bigskip
\bigskip
\bigskip

\begin{figure}[bthp]
\begin{center}
$\begin{array}{c}\\
\includegraphics[height=5.5cm,keepaspectratio]{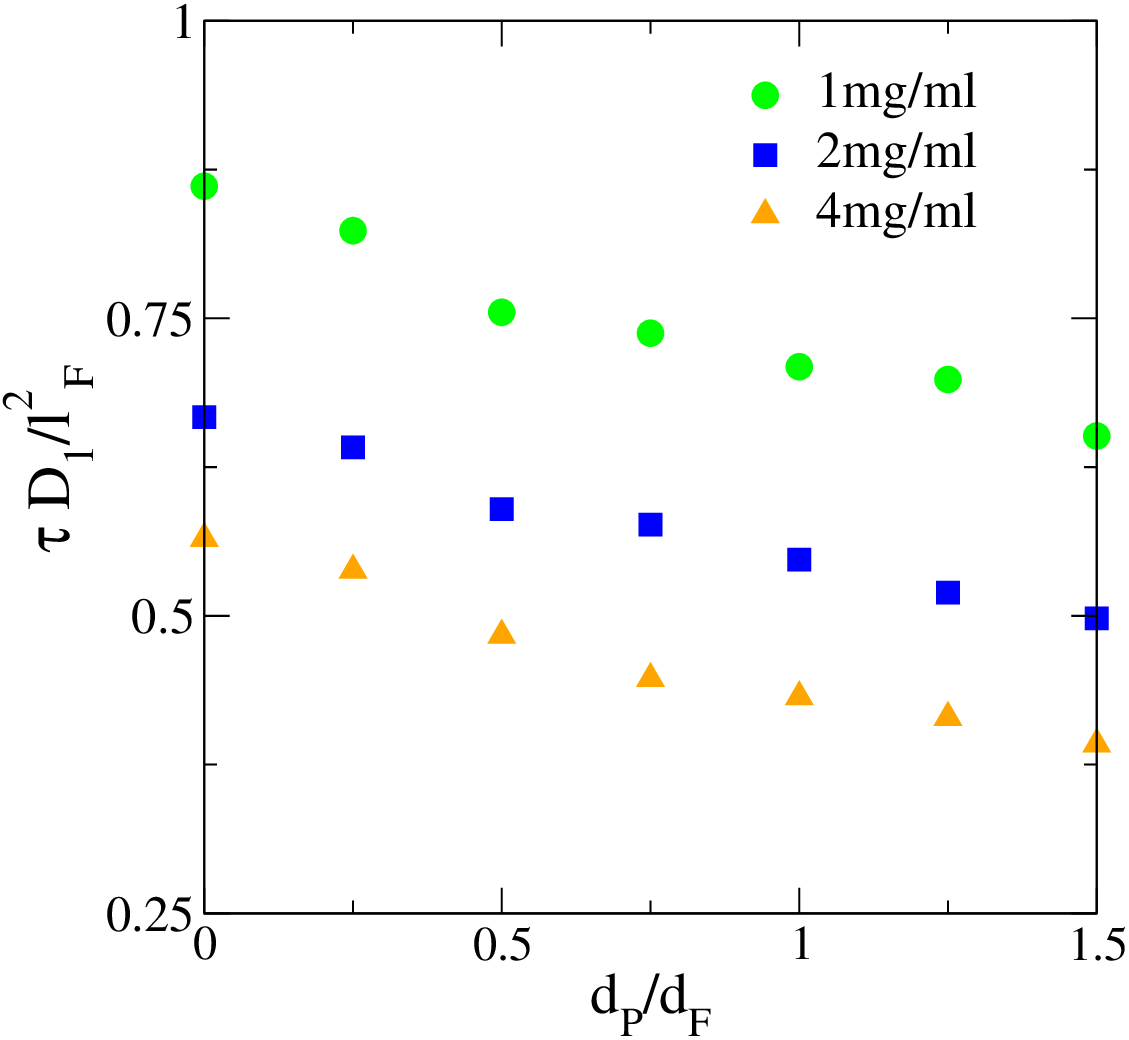} \\
\end{array}$
\end{center}
\caption{Jiao and Torquato}
\label{fig_tau}
\end{figure}
\bigskip
\bigskip
\bigskip
\bigskip

\begin{figure}[bthp]
\begin{center}
$\begin{array}{c@{\hspace{0.6cm}}c@{\hspace{0.6cm}}c}\\
\includegraphics[height=4.5cm,keepaspectratio]{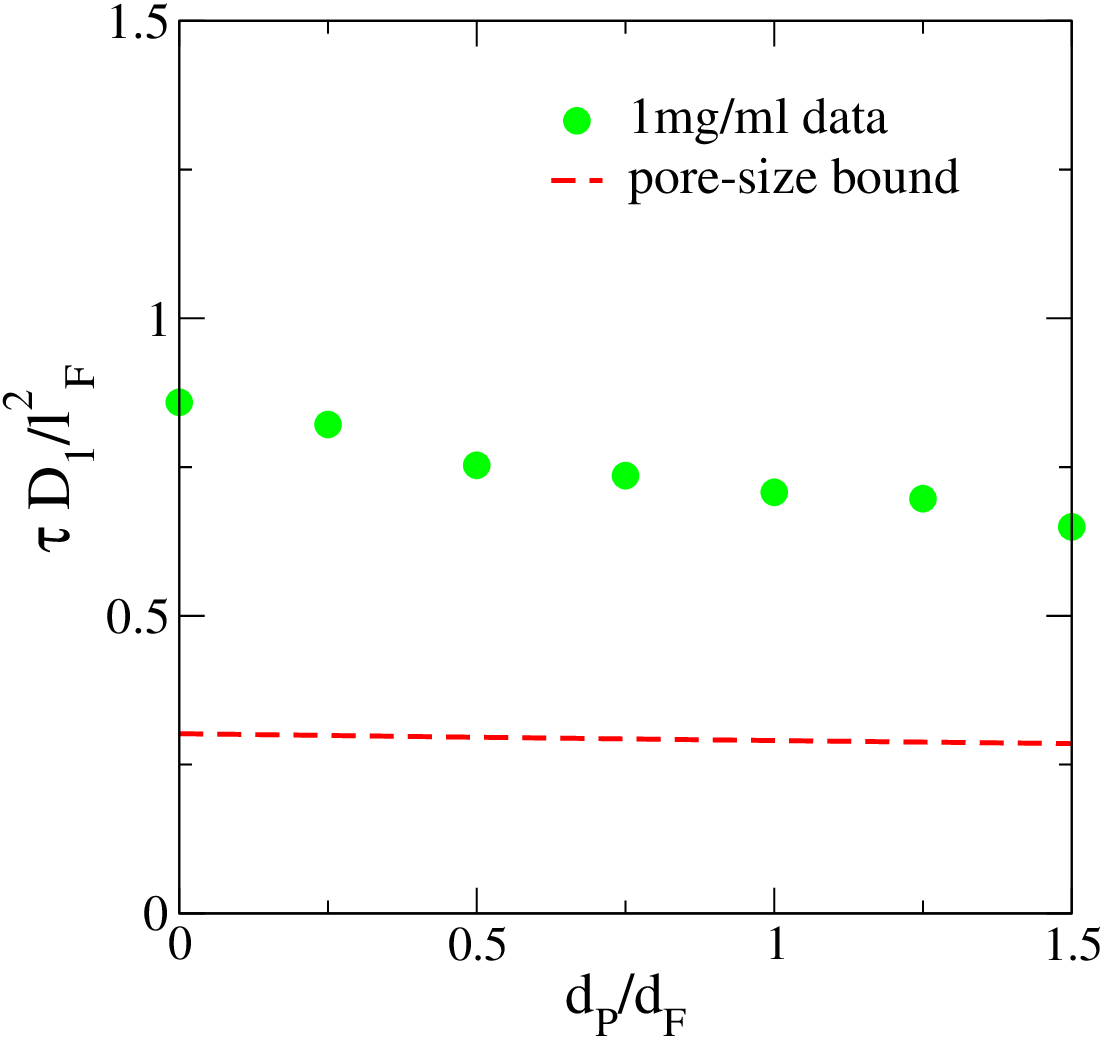} &
\includegraphics[height=4.5cm,keepaspectratio]{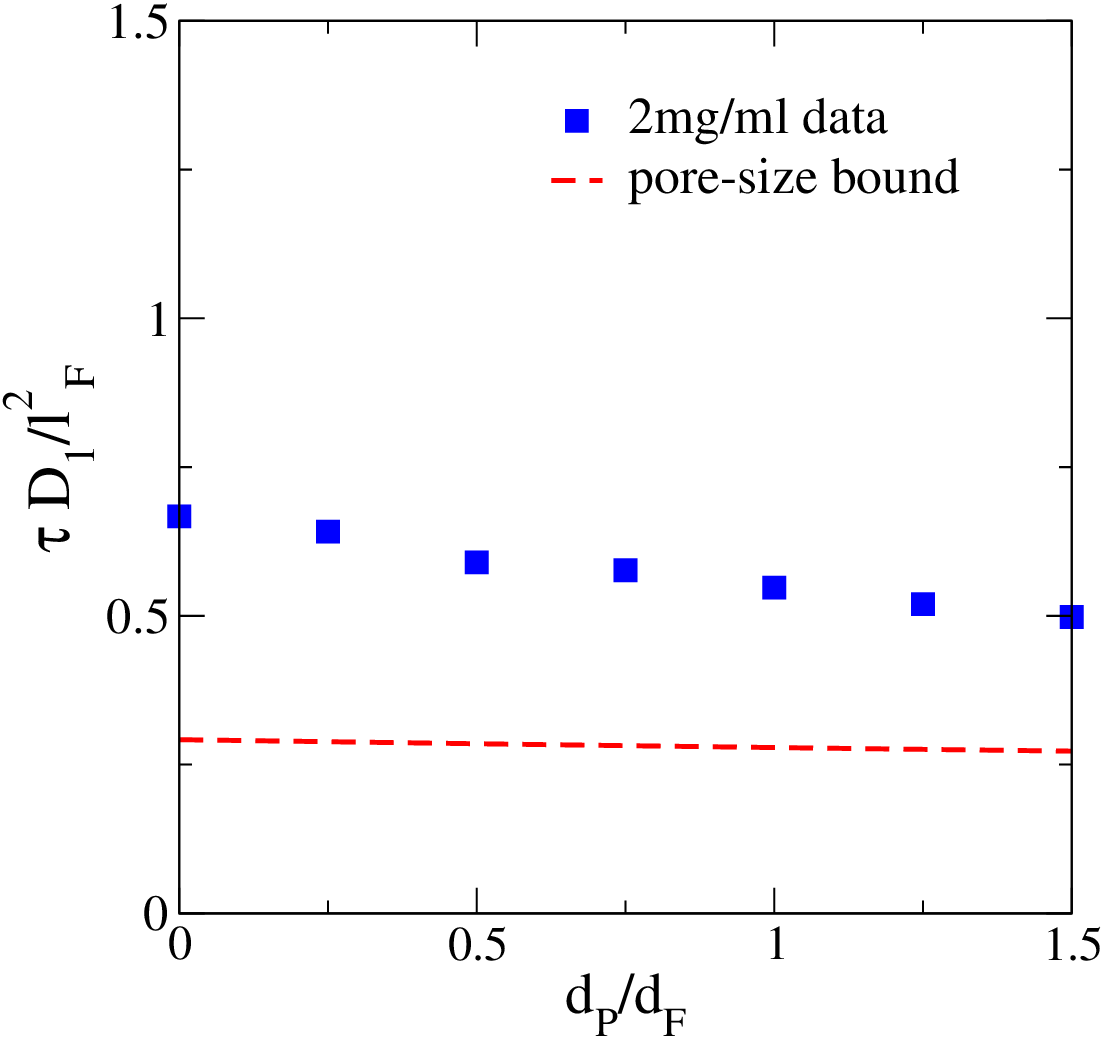} &
\includegraphics[height=4.5cm,keepaspectratio]{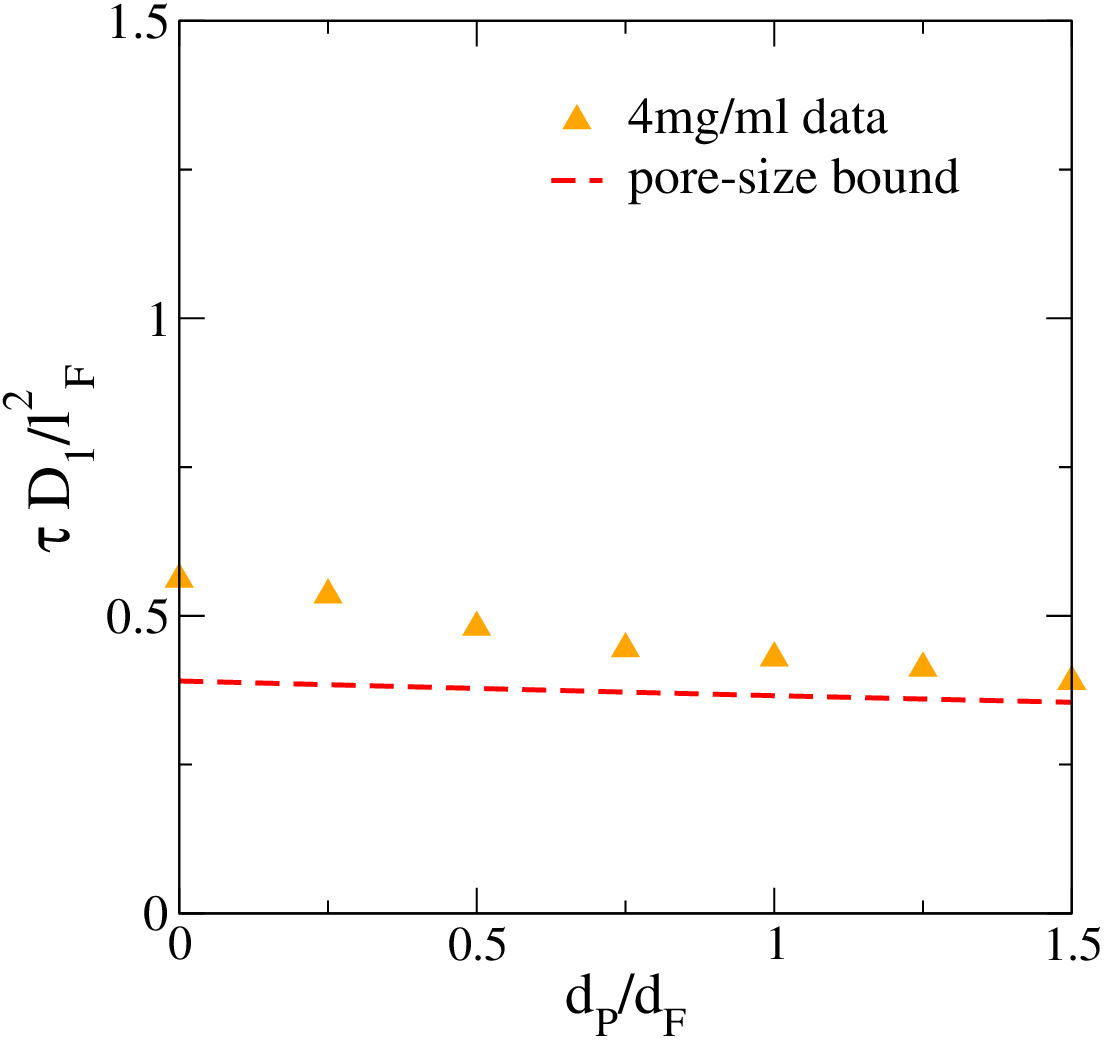} \\
\mbox{\bf (a)} & \mbox{\bf (b)} & \mbox{\bf (c)}
\end{array}$
\end{center}
\caption{Jiao and Torquato}
\label{fig_tau_bd}
\end{figure}
\bigskip
\bigskip
\bigskip
\bigskip

\begin{figure}[bthp]
\begin{center}
$\begin{array}{c}\\
\includegraphics[height=5.5cm,keepaspectratio]{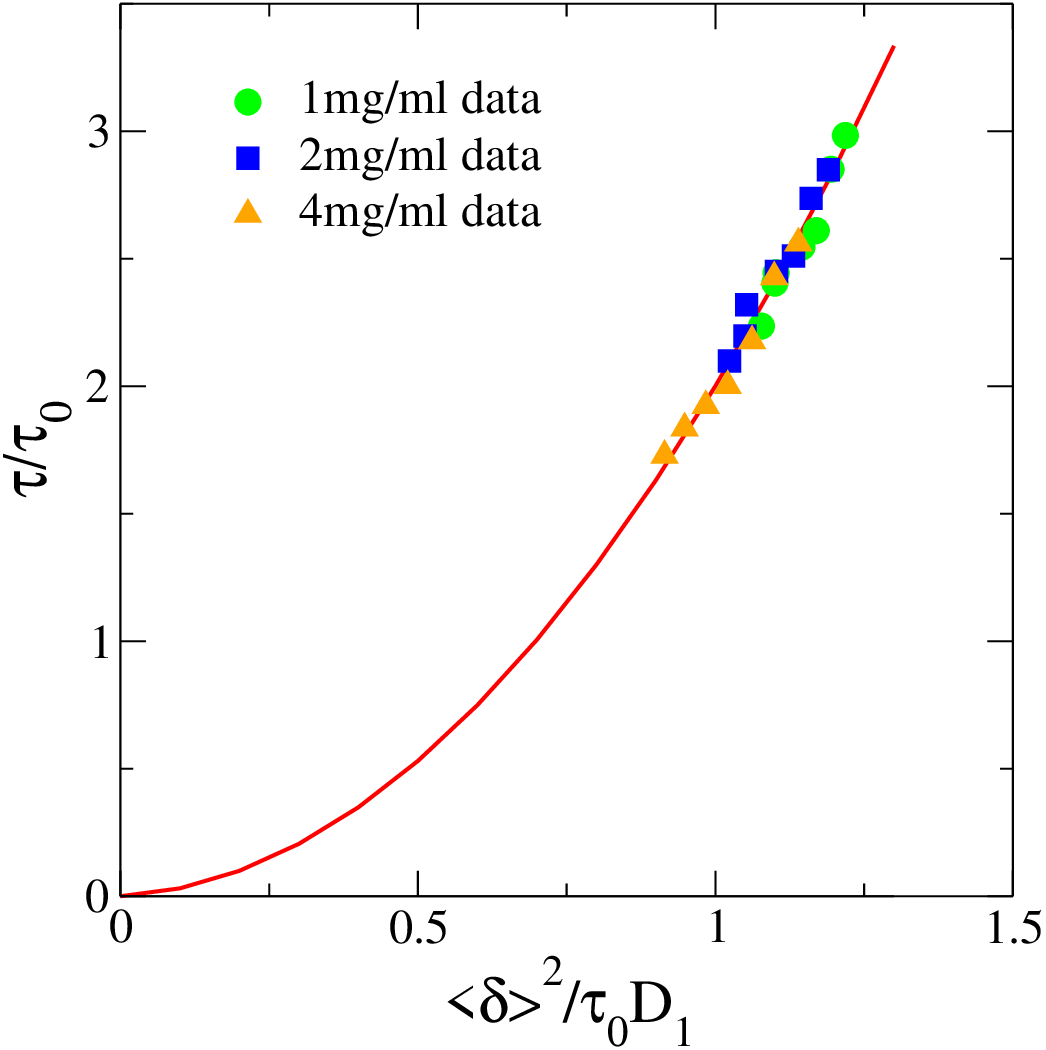} \\
\end{array}$
\end{center}
\caption{Jiao and Torquato}
\label{fig_tau_univ}
\end{figure}
\bigskip
\bigskip
\bigskip
\bigskip

\begin{figure}[bthp]
\begin{center}
$\begin{array}{c}\\
\includegraphics[height=5.5cm,keepaspectratio]{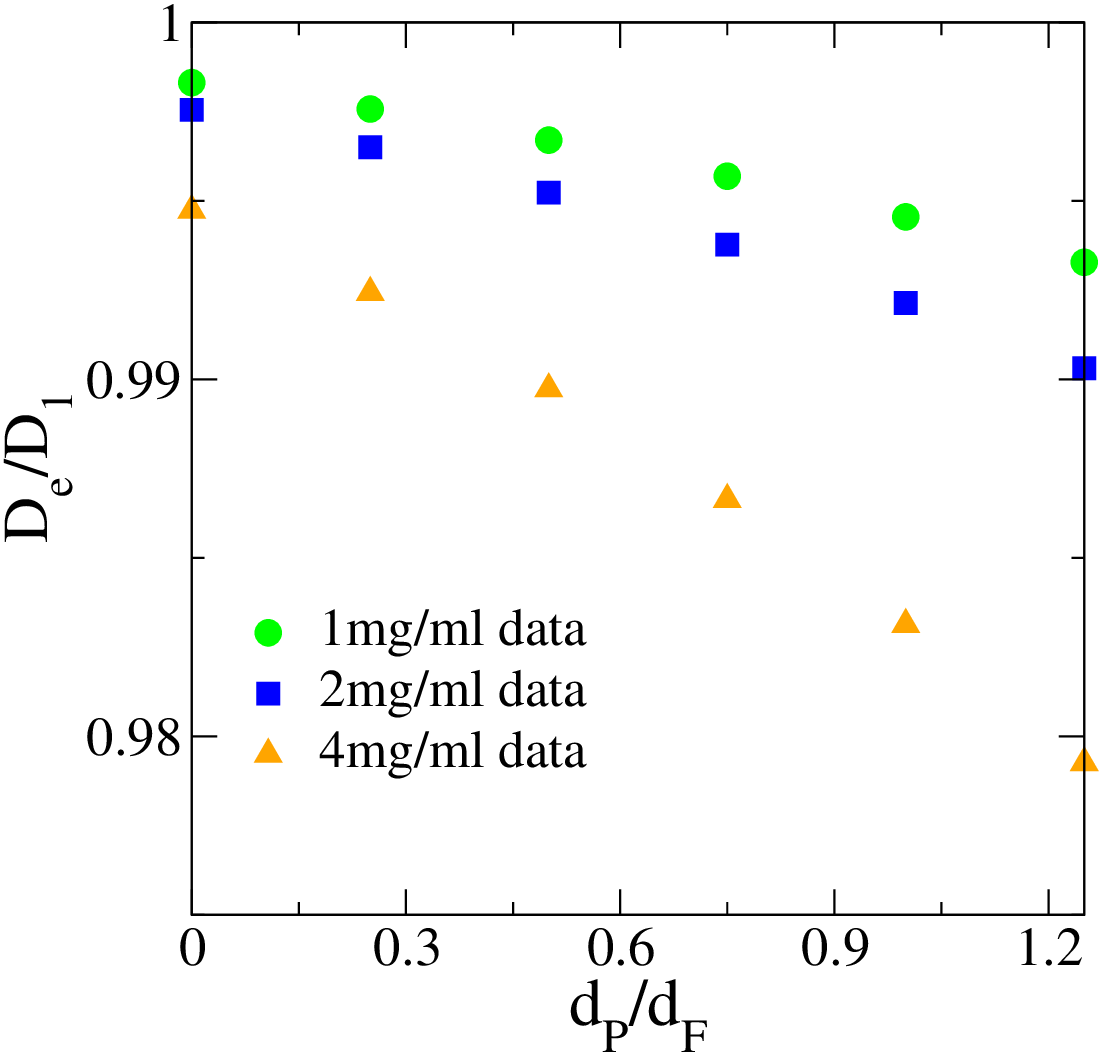} \\
\end{array}$
\end{center}
\caption{Jiao and Torquato}
\label{fig_De}
\end{figure}
\bigskip
\bigskip
\bigskip
\bigskip

\begin{figure}[bthp]
\begin{center}
$\begin{array}{c}\\
\includegraphics[height=6.5cm,keepaspectratio]{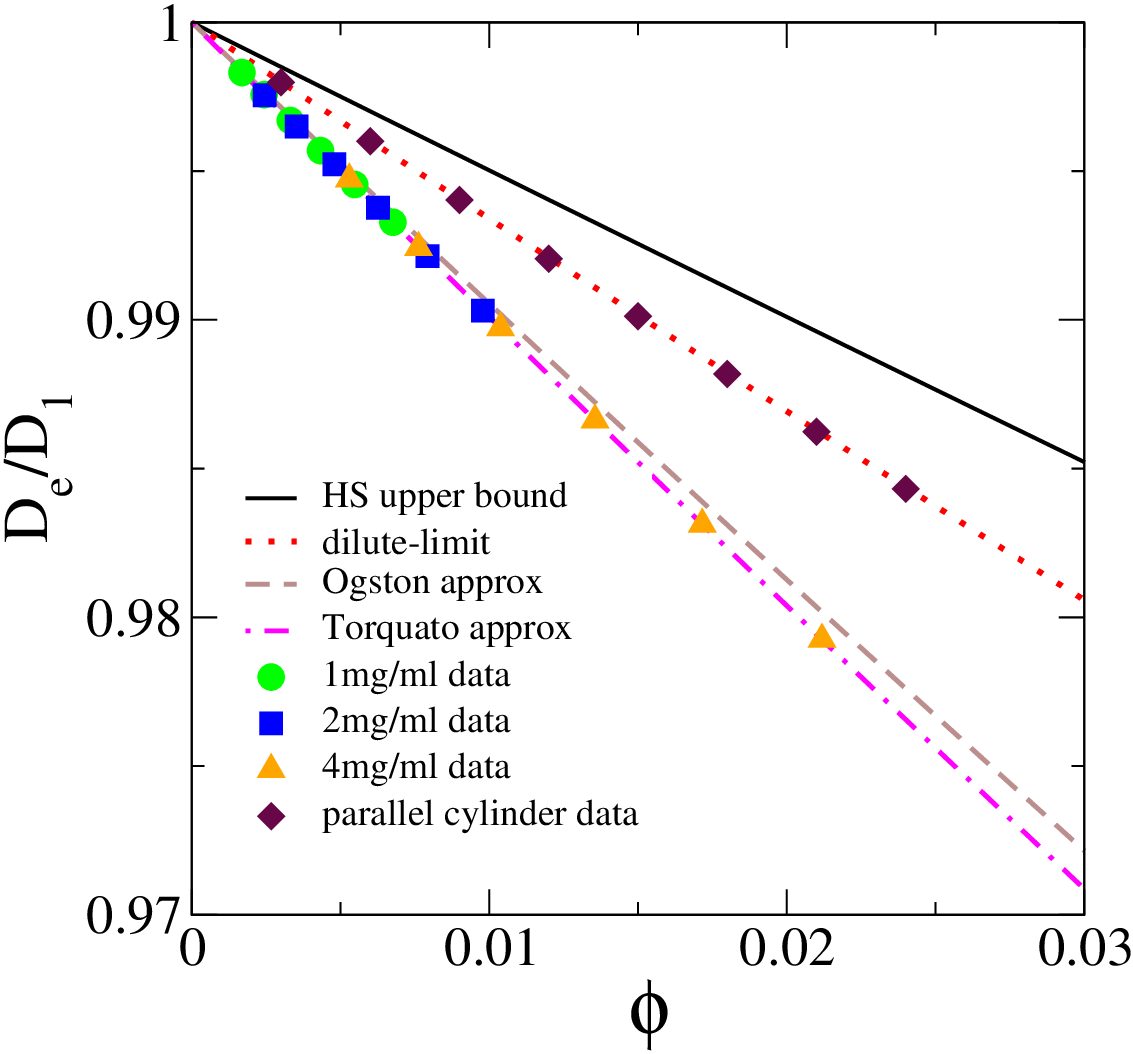} \\
\end{array}$
\end{center}
\caption{Jiao and Torquato}
\label{fig_De_bd}
\end{figure}



\end{document}